\documentclass[twocolumn]{IEEEtran}
\normalsize
\usepackage[usenames,dvipsnames]{xcolor}
\usepackage[cmex10]{amsmath}
\usepackage{amssymb}
\usepackage{graphicx}
\usepackage{epstopdf}
\usepackage[noadjust]{cite}
\usepackage{etoolbox}
\usepackage{filecontents}
\usepackage{tikz}
\usetikzlibrary{shapes, arrows, calc}
\def\rightantenna{%
	-- +(1.5mm,0mm)
    -- +(1.5mm,1mm) -- +(2.5mm,2.5mm) -- +(0.5mm,2.5mm) -- +(1.5mm,1mm)
}
\def\leftantenna{%
	-- +(-1.9mm,0mm)	
    -- +(-1.9mm,1.0mm) -- +(-2.9mm,2.5mm) -- +(-0.9mm,2.5mm) -- +(-1.9mm,1.0mm)
}
\tikzset{
    right angle quadrant/.code={
        \pgfmathsetmacro\quadranta{{1,1,-1,-1}[#1-1]}     % Arrays for selecting quadrant
        \pgfmathsetmacro\quadrantb{{1,-1,-1,1}[#1-1]}},
    right angle quadrant=1, % Make sure it is set, even if not called explicitly
    right angle length/.code={\def\rightanglelength{#1}},   % Length of symbol
    right angle length=1ex, % Make sure it is set...
    right angle symbol/.style n args={3}{
        insert path={
            let \p0 = ($(#1)!(#3)!(#2)$) in     % Intersection
                let \p1 = ($(\p0)!\quadranta*\rightanglelength!(#3)$), % Point on base line
                \p2 = ($(\p0)!\quadrantb*\rightanglelength!(#2)$) in % Point on perpendicular line
                let \p3 = ($(\p1)+(\p2)-(\p0)$) in  % Corner point of symbol
            (\p1) -- (\p3) -- (\p2)
        }
    }
}
\newcounter{MYtempeqncnt}
\begin{document}
\title{User Selection in MIMO Interfering Broadcast Channels}
\author{Gaurav~Gupta and A.K.~Chaturvedi,~\IEEEmembership{Senior~Member,~IEEE}
\thanks{This work was supported by BSNL-IITK Telecom Centre of Excellence at IIT Kanpur.}

\thanks{The work reported in this paper was done at the Department of Electrical Engineering, Indian Institute of Technology Kanpur, India (email: gaurav71531@gmail.com; akc@iitk.ac.in).}}
\maketitle
%
%--------------------------------------------------------------------------------
%-----------------------------------ABSTRACT-------------------------------------
%--------------------------------------------------------------------------------
%
\begin{abstract}
Interference alignment aims to achieve maximum degrees of freedom in an interference system. For achieving Interference alignment in interfering broadcast systems a closed-form solution is proposed in \cite{tang} which is an extension of the grouping scheme in \cite{shin}. In a downlink scenario where there are a large number of users, the base station is required to select a subset of users such that the sum rate is maximized. To search for the optimal user subset using brute-force approach is computationally exhaustive because of the large number of possible user subset combinations. We propose a user selection algorithm achieving sum rate close to that of optimal solution. The algorithm employs coordinate ascent approach and exploits orthogonality between the desired signal space and the interference channel space in the reciprocal system to select the user at each step. For the sake of completeness, we have also extended the sum rate approach based algorithm to Interfering broadcast channel. The complexity of both these algorithms is shown to be linear with respect to the total number of users as compared to exponential in brute-force search.
\end{abstract}
%
%--------------------------------------------------------------------------------
%-----------------------------SEC-I : INTRODUCTION-------------------------------
%--------------------------------------------------------------------------------
%
\section{Introduction}
\label{sec:intro}
\IEEEPARstart{M}{ultiuser} systems with interference from multiple transmitters has attracted a lot of attention in recent times. The authors of \cite{jafar} proposed an Interference Alignment (IA) scheme to achieve maximum degrees of freedom (dof) in a $K$-transmitter and $K$-receiver (or $K$-user) time-varying interference channel (IFC) with single antenna at each transmitter and receiver. For a system having multiple antennas and identical antenna configuration at each node, IA can be achieved with constant channels also. However, the closed-form solution for the precoder to achieve IA is known only for the three-user IFC with global channel knowledge at each node. Since this closed-form solution does not take sum rate maximization into account, in \cite{sung} the precoder is optimized to jointly achieve IA and sum rate maximization. For the general case $(K \ge 3)$, by using reciprocity of the network two iterative algorithms have been proposed in \cite{gomadan} which require only local channel knowledge at each node.

Now consider a cellular network, referred to as Interfering Broadcast Channel (IFBC), in which each base station (BS) supports multiple users. The IFC supports single user in each cell and hence there is only inter-cell interference (ICI) while IFBC can support multiple users, therefore there is inter-user interference (IUI) as well as ICI at each receiver. In order to deal with these interferences we need to design the transmitter and receiver beamformer. This was addressed in \cite{park} by proposing a coordinated Zero-forcing (ZF) scheme to mitigate both IUI and ICI in a multiple-input single-output (MISO) IFBC. In \cite{kim} the coordinated ZF scheme was extended to MIMO-IFBC by considering multiple antennas at each receiver. By building upon the notion of IA in \cite{jafar, motahari}, a new precoding scheme called subspace IA was introduced in \cite{suh}. The scheme is based on alignment of ICI and IUI into a multi-dimensional subspace instead of one dimension. Subsequently the authors of \cite{ho} developed an IA technique for a downlink cellular system which requires feedback only within its cell. The scheme offered substantial advantages when interference from a dominant interferer is significantly stronger than the remaining interference. 

To avoid an iterative procedure and to achieve optimal dof, in \cite{shin} a grouping method was proposed for the two cell and two user MIMO-IFBC having different number of antennas at the transmitter and the receiver. The key idea behind the grouping scheme is to cooperatively construct the receive beamformer so as to align the effective ICI channel. This helps the BS to treat these ICI channels as one effective ICI channel and accordingly construct the transmit beamformer. The transmit beamformer lies in the space orthogonal to the space spanned by IUI channels' and the effective ICI channel to completely eliminate the interference received at the user. In \cite{tang} the grouping scheme was extended to more than two cells and more than two users in each cell. This extension reduces the cost in terms of number of transmit antennas required to achieve the same dof.

While the user selection problem has been addressed in the literature for the IFC case, the same is not true for the IFBC. A user selection algorithm for the three-user MIMO-IFC was proposed in \cite{inkyu}. The algorithm used a closed-form solution to design the precoding matrices \cite{jafar} and Minimum Mean Squared Error (MMSE) receive beamformer. To utilize the multiuser diversity, in \cite{inkyu} the users at each step are selected by employing coordinate ascent approach \cite{ascent}. An opportunistic user selection algorithm for a three-user MIMO-IFC was proposed in \cite{jung_jour}. The algorithm selects the user, the interference channels' of which have maximum alignment with each other. Random beamforming is performed at each transmitter and then post-processing is performed only at the users selected by the algorithm.

In this paper, to improve the achievable sum rate, we address the problem of user selection in IFBC. The extended grouping method has been used to achieve IA. The user selection in IFBC with IA is complicated by the fact that changing the effective channel of any user in a given cell has effects on the effective channel of the remaining users in its own cell as well as on all the users in rest of the cells. In addition, because of grouping, the receive beamforming matrix of each user has a special structure which relates its effective channel to the interference channels' from its neighboring BS. We propose a low complexity user selection algorithm with the goal of maximizing the sum rate of the system. The algorithm exploits orthogonality between desired signal space and interference channel space in the reciprocal system to select a user. For the sake of completeness, we will also extend the sum rate based algorithm in \cite{inkyu} to IFBC. Both algorithms use coordinate ascent approach \cite{ascent} to update the user subset iteratively and are shown to have computation complexity linear in the number of users in each cell and achieve sum rate close to that achieved by the optimal user subset. 
%
%--------------------------------------------------------------------------------
%-----------------------------SEC-II : SYSTEM MODEL------------------------------
%--------------------------------------------------------------------------------
%
\section{System Model and Background}
\label{sec:sys_model}

\begin{figure}
\centering
\begin{tikzpicture}[scale = 1.4]

\draw (0.2,0) rectangle (1,1) ;
\node [anchor = west, color=blue] (BS3) at (0.3,0.5) {$\textbf{BS-3}$};
\draw (1,0.1) \rightantenna;
\draw (1,0.6) \rightantenna;
\draw (1,0.9) \rightantenna;
\draw[dotted, line width=2pt] (1.1,0.4) -- (1.1,0.6);
\node [anchor = west] (BS3_V) at (-1,0.5) {$[\textbf{V}_{1}^{[3]},\textbf{V}_{2}^{[3]}]$};

\draw (0.2,2) rectangle (1,3) ;
\node [anchor = west, color=blue] (BS2) at (0.3,2.5) {$\textbf{BS-2}$};
\draw (1,2.1) \rightantenna;
\draw (1,2.6) \rightantenna;
\draw (1,2.9) \rightantenna;
\draw[dotted, line width=2pt] (1.1,2.4) -- (1.1,2.6);
\node [anchor = west] (BS2_V) at (-1,2.5) {$[\textbf{V}_{1}^{[2]},\textbf{V}_{2}^{[2]}]$};

\draw (0.2,4) rectangle (1,5) ;
\node [anchor = west, color=blue] (BS1) at (0.3,4.5) {$\textbf{BS-1}$};
\draw (1,4.1) \rightantenna;
\draw (1,4.6) \rightantenna;
\draw (1,4.9) \rightantenna;
\draw[dotted, line width=2pt] (1.1,4.4) -- (1.1,4.6);
\node [anchor = west] (BS3_V) at (-1,4.5) {$[{\color{ForestGreen}\textbf{V}_{1}^{[1]}},{\color{Red}\textbf{V}_{2}^{[1]}}]$};

\draw[ ->, color=ForestGreen, line width=0.8pt] (1.3,4.5) -- (3.6,4.8) node[pos = 0.5, sloped, above] {{\color{black}desired signal}};
\draw[loosely dashed,->, color=red, line width=0.8pt] (1.3,4.5) -- (3.6,4.2);
\draw[dotted,->, color=red, line width=0.8pt] (1.3,4.5) -- (3.6,0.2);
\draw[dotted,->, color=red, line width=0.8pt] (1.3,4.5) -- (3.6,0.8);
\draw[dotted,->, color=red, line width=0.8pt] (1.3,4.5) -- (3.6,2.2);
\draw[dotted,->, color=red, line width=0.8pt] (1.3,4.5) -- (3.6,2.8);

\draw (4,0) rectangle (4.7,0.4) ;
\node [anchor = west, color=blue] (User23) at (4,0.2) {$\textbf{User}_{2}^{3}$};
\draw (4,0) \leftantenna;
\draw (4,0.3) \leftantenna;
\draw[dotted, line width = 1.5pt] (3.95,0.05) -- (3.95,0.25);
\node [anchor = west, color=red] (U23) at (4.7,0.2) {$\textbf{U}_{2}^{[3]}$};

\draw (4,0.6) rectangle (4.7,1) ;
\node [anchor = west, color=blue] (User13) at (4,0.8) {$\textbf{User}_{1}^{3}$};
\draw (4,0.6) \leftantenna;
\draw (4,0.9) \leftantenna;
\draw[dotted, line width = 1.5pt] (3.95,0.65) -- (3.95,0.85);
\node [anchor = west, color=red] (U13) at (4.7,0.8) {$\textbf{U}_{1}^{[3]}$};

\draw (4,2) rectangle (4.7,2.4) ;
\node [anchor = west, color=blue] (User22) at (4,2.2) {$\textbf{User}_{2}^{2}$};
\draw (4,2) \leftantenna;
\draw (4,2.3) \leftantenna;
\draw[dotted, line width = 1.5pt] (3.95,2.05) -- (3.95,2.25);
\node [anchor = west, color=red] (U22) at (4.7,2.2) {$\textbf{U}_{2}^{[2]}$};

\draw (4,2.6) rectangle (4.7,3) ;
\node [anchor = west, color=blue] (User12) at (4,2.8) {$\textbf{User}_{1}^{2}$};
\draw (4,2.6) \leftantenna;
\draw (4,2.9) \leftantenna;
\draw[dotted, line width = 1.5pt] (3.95,2.65) -- (3.95,2.85);
\node [anchor = west, color=red] (U12) at (4.7,2.8) {$\textbf{U}_{1}^{[2]}$};

\draw (4,4) rectangle (4.7,4.4) ;
\node [anchor = west, color=blue] (User21) at (4,4.2) {$\textbf{User}_{2}^{1}$};
\draw (4,4) \leftantenna;
\draw (4,4.3) \leftantenna;
\draw[dotted, line width = 1.5pt] (3.95,4.05) -- (3.95,4.25);
\node [anchor = west, color=red] (U21) at (4.7,4.2) {$\textbf{U}_{2}^{[1]}$};

\draw (4,4.6) rectangle (4.7,5) ;
\node [anchor = west, color=blue] (User11) at (4,4.8) {$\textbf{User}_{1}^{1}$};
\draw (4,4.6) \leftantenna;
\draw (4,4.9) \leftantenna;
\draw[dotted, line width = 1.5pt] (3.95,4.65) -- (3.95,4.85);
\node [anchor = west, color=ForestGreen] (U11) at (4.7,4.8) {$\textbf{U}_{1}^{[1]}$};

\node [anchor = west] (IUI) at (2.8,3.8) {inter-user interference};
\draw[->] (2.8,3.8) -- (2.6,4.3);
\draw[dotted, line width=0.8pt, color = red] (2,3.4) ellipse (0.3 and 0.7);
\node [anchor = west] (ICI) at (1,1.5) {inter-cell interference};
\draw[->] (1.5,1.6) -- (2,2.7);

\end{tikzpicture}
\caption{MIMO-IFBC with $L = 3$ and $K = 2$ in each cell where the BS-$1$ is shown to be generating IUI and ICI for the users in its own cell and neighboring cells respectively.}
\label{fig:IFBC_sys}
\end{figure}
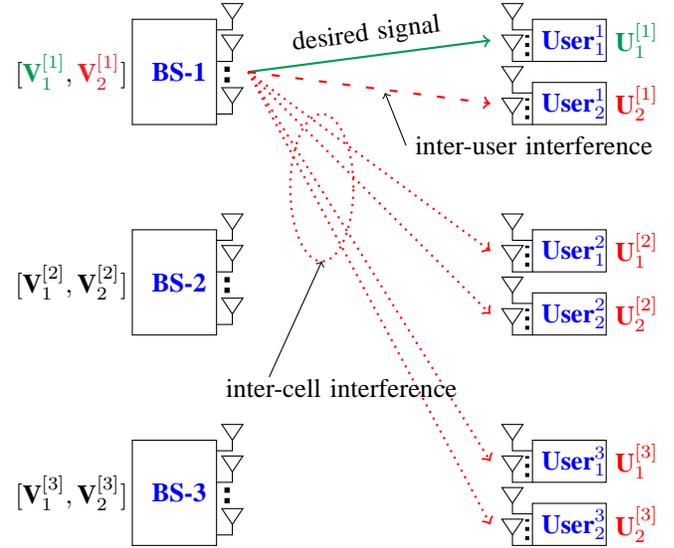

We consider a MIMO-IFBC downlink cellular system with $L$ cells such that each cell has one BS and supports $K$ users. We assume that each transmitting node (BS) is equipped with $M$ antennas and each receiving node is equipped with $N$ antennas, where $M > N$. For example, in \figurename\,\ref{fig:IFBC_sys} we have shown a MIMO-IFBC cellular network with three cells and each BS supporting two users. We also assume that each BS tries to convey $d_{s}$ data streams to each user such that $d_{s} \le \text{min}\left(M, N\right) = N$. The transmit signal intended for the $k$th user in the $l$th cell is given by

\begin{equation}
{\bf x}_{k}^{[l]} = \sum\limits_{i = 1}^{d_{s}}{\bf v}_{k,i}^{[l]}s_{k,i}^{[l]} = {\bf V}_{k}^{[l]}{\bf s}_{k}^{[l]}
\label{eqn:trans_sig}
\end{equation}

\noindent where $s_{k,i}^{[l]}$ is the $i$th symbol precoded using the linear beamforming vector ${\bf v}_{k}^{[l]} \in \mathbb{C}^{M\times 1}$ with $||{\bf v}_{k}^{[l]}|| = 1$. The transmit power constraint at the $l$th BS is $\mathbb{E}\left\{\sum\nolimits_{k = 1}^{K}||{\bf x}_{k}^{[l]}||^2\right\} \le P_{l}$. 
The $M\times d_{s}$ transmit beamforming matrix is denoted as ${\bf V}_{k}^{[l]} = [{\bf v}_{k,1}^{[l]},{\bf v}_{k,2}^{[l]},...,{\bf v}_{k,d_{s}}^{[l]}]$ and the corresponding $d_{s}\times 1$ symbol vector is denoted as ${\bf s}_{k}^{[l]} = [s_{k,1}^{[l]},s_{k,2}^{[l]},...,s_{k,d_{s}}^{[l]}]^{T}$. The received signal at the $k$th user in the $l$th cell is given by

\begin{eqnarray}
{\bf y}_{k}^{[l]} &=& \sum\limits_{j = 1}^{L}{\bf H}_{k}^{[l,j]}\sum\limits_{i = 1}^{K}{\bf x}_{i}^{[j]} \nonumber \\
&=& \underbrace{{\bf H}_{k}^{[l,l]}{\bf V}_{k}^{[l]}{\bf s}_{k}^{[l]}}_{\text{desired signal}} + \underbrace{\sum\limits_{i = 1, i \neq k}^{K}{\bf H}_{k}^{[l,l]}{\bf V}_{i}^{[l]}{\bf s}_{i}^{[l]}}_{\text{inter-user interference}} \nonumber \\
&& {+}\: \underbrace{\sum\limits_{j = 1, j \neq l}^{L}\sum\limits_{i = 1}^{K}{\bf H}_{k}^{[l,j]}{\bf V}_{i}^{[j]}{\bf s}_{i}^{[j]}}_{\text{inter-cell interference}} {+}\: {\bf n}_{k}^{[l]}
\label{eqn:rec_sig}
\end{eqnarray}

\noindent where ${\bf H}_{k}^{[l,j]} \in \mathbb{C}^{N\times M}$ is the channel matrix from the $j$th BS to the $k$th user in the $l$th cell, each entry of which is independently and identically distributed (i.i.d.) circular symmetric complex Gaussian random variable with unit variance. The channel is assumed to be slow-varying flat fading. Each entry of the $N\times 1$ Additive White Gaussian Noise (AWGN) vector ${\bf n}_{k}^{[l]}$ is assumed to be i.i.d. complex random variable with variance $\sigma^2$. Each user performs receive beamforming operation to take care of the interference received. The $k$th user in the $l$th cell detects the received signal as

\begin{align}
\tilde{{\bf y}}_{k}^{[l]} &= {\bf U}_{k}^{[l]H}{\bf H}_{k}^{[l,l]}{\bf V}_{k}^{[l]}{\bf s}_{k}^{[l]} \nonumber \\
&{+}\: {\bf U}_{k}^{[l]H}(\sum\limits_{i = 1, i \neq k}^{K}{\bf H}_{k}^{[l,l]}{\bf V}_{i}^{[l]}{\bf s}_{i}^{[l]} + \sum\limits_{j = 1, j \neq l}^{L}\sum\limits_{i = 1}^{K}{\bf H}_{k}^{[l,j]}{\bf V}_{i}^{[j]}{\bf s}_{i}^{[j]}) \nonumber \\
&{+}\: \tilde{{\bf n}}_{k}^{[l]}
\label{eqn:dec_sig}
\end{align}

\noindent where ${\bf U}_{k}^{[l]}$ is the $d_{s}\times N$ receive beamforming matrix for the $k$th user in the $l$th cell and $\tilde{{\bf n}}_{k}^{[l]} = {\bf U}_{k}^{[l]H}{\bf n}_{k}^{[l]}$.

%
%--------------------------------------------------------------------------------
%----------------------------FIG-I : MIMO-IFBC----------------------------------- %--------------------------------------------------------------------------------
%
%--------------------------------------------------------------------------------

\subsection{Interference Cancellation}

For efficient detection of the desired signal, the receiver is required to project the received signal onto orthogonal space of the interference received. The following feasibility conditions \cite{tang} need to be satisfied.

\begin{IEEEeqnarray}{rCl}
&&{\bf U}_{k}^{[l]H}{\bf H}_{k}^{[l,l]}{\bf V}_{i}^{[l]} = 0, \quad \forall i \neq k  
\label{eqn:rec_reqr_1}\\
&&{\bf U}_{k}^{[l]H}{\bf H}_{k}^{[l,j]}{\bf V}_{m}^{[j]} = 0, \quad \forall m = 1,...,K \text{ and } \forall j \neq l  \label{eqn:rec_reqr_2}\\
&&\text{rank}\left({\bf U}_{k}^{[l]H}{\bf H}_{k}^{[l,l]}{\bf V}_{k}^{[l]} \right) = d_{s}
\label{eqn:rec_reqr_3}
\end{IEEEeqnarray}

The desired signal can now be interpreted as received through a $d_{s}\times d_{s}$ effective channel matrix

\begin{equation}
\bar{{\bf H}}_{k}^{[l,l]} =  {\bf U}_{k}^{[l]H}{\bf H}_{k}^{[l,l]}{\bf V}_{k}^{[l]}
\label{eqn:eff_channel}
\end{equation}

\begin{figure*}[!t]
 \normalsize
 \setcounter{MYtempeqncnt}{\value{equation}}
 \setcounter{equation}{10}
 \begin{equation}
  \begin{bmatrix}
   {\bf I}_{M} & -{\bf H}_{1}^{[l+1,l]H} & {\bf 0} &\hdots & {\bf 0} \\
   {\bf I}_{M} & {\bf 0} & -{\bf H}_{2}^{[l+1,l]H} & \hdots & {\bf 0} \\
   \vdots & \vdots & \vdots & \ddots & \vdots \\
   {\bf I}_{M} & {\bf 0} & {\bf 0} &\hdots & -{\bf H}_{K}^{[l+1,l]H}
  \end{bmatrix}
  \begin{bmatrix}
   {\bf G}_{l} \\
   {\bf U}_{1}^{[l+1]} \\
   {\bf U}_{2}^{[l+1]} \\
   \vdots \\
   {\bf U}_{K}^{[l+1]}
  \end{bmatrix} = {\bf F}_{l}{\bf X}_{l} = {\bf 0}
  \label{eqn:U_mat}
 \end{equation}
 \hrulefill
 \begin{equation}
  {\bf V}_{k}^{[l]} \subset \text{null}([
   \underbrace{{\bf G}_{l}}_{\text{effective interference channels}} \quad
   \underbrace{({\bf U}_{t(t=1,...,K)}^{[s(s\neq l,\neq l+1)]H}{\bf H}_{t(t=1,...,K)}^{[s(s\neq l,\neq l+1), l]})^{H}}_{\text{effective ICI channels}} 
   \quad
   \underbrace{({\bf U}_{t(t=1,...,K, \neq k)}^{[l]H}{\bf H}_{t(t=1,...,K, \neq k)}^{[l,l]})^{H}}_{\text{effective IUI channels}}
   ]^{H})
  \label{eqn:V_mat}
 \end{equation}
 \hrulefill
 \vspace*{-5pt}
\end{figure*}

The effective noise at the $k$th user in the $l$th cell, $\tilde{{\bf n}}_{k}^{[l]}$ is colored and hence a pre-whitening filter is required. A pre-whitening filter of the form ${\bf W}_{k}^{[l]} = ({\bf U}_{k}^{[l]H}{\bf U}_{k}^{[l]})^{-1/2}$ will be used at the user to detect the received $d_{s}$ symbols independently. The received signal after pre-whitening filter can be expressed as

\begin{equation}
\setcounter{equation}{7}
 \bar{{\bf y}}_{k}^{[l]} = {\bf W}_{k}^{[l]}\bar{{\bf H}}_{k}^{[l,l]}{\bf s}_{k}^{[l]} + \bar{{\bf n}}_{k}^{[l]}
 \setcounter{equation}{8}
 \label{eqn:white_eff}
\end{equation}

\noindent where $\bar{{\bf n}}_{k}^{[l]} = {\bf W}_{k}^{[l]}\tilde{{\bf n}}_{k}^{[l]}$ and hence $\mathbb{E}\{\bar{{\bf n}}_{k}^{[l]}\bar{{\bf n}}_{k}^{[l]H} \} = {\bf I}_{d_{s}}$. The sum rate achieved can be written as

\begin{IEEEeqnarray}{rCl}
 \mathcal{R} &=& \sum\limits_{l = 1}^{L}\sum\limits_{k = 1}^{K}\mathcal{R}_{k}^{[l]} \nonumber\\ 
 &=& \sum\limits_{l = 1}^{L}\max\limits_{\left\{{\bf Q}_{k}^{[l]}:{\bf Q}_{k}^{[l]} \ge 0, \sum\limits_{k = 1}^{K}\text{tr}({\bf Q}_{k}^{[l]}) \le P_{l}\right\}} \nonumber \\
 &&\quad\sum\limits_{k = 1}^{K} \text{log}_{2}\left|{\bf I}_{d_{s}} + \frac{1}{\sigma^{2}}{\bf W}_{k}^{[l]}\bar{{\bf H}}_{k}^{[l,l]}{\bf Q}_{k}^{[l]}\bar{{\bf H}}_{k}^{[l,l]H}{\bf W}_{k}^{[l]H}\right|
 \label{eqn:water_fill}
\end{IEEEeqnarray}

\noindent where ${\bf Q}_{k}^{[l]} = \mathbb{E}\{{\bf x}_{k}^{[l]}{\bf x}_{k}^{[l]H}\}$ is the input covariance matrix of the $k$th user in the $l$th cell. The solution to the RHS of (\ref{eqn:water_fill}) can be found by using the well known water-filling algorithm with the power constraint $P_{l}$ for all $l$.

\subsection{Extended Grouping Scheme}

The extended grouping scheme \cite{tang} is a generalization of the non-iterative grouping scheme \cite{shin} for a multi-cell and multi-user system with complete suppression of interference. We will briefly touch the basic aspects of the extended grouping scheme.

The grouping of users is achieved by appropriately designing the receiver beamforming matrices ${\bf U}_{k}^{[l]}$ for all the users in any given cell. The users in the cell next\footnote{here next refers to cyclic next, for example in our $L$-cell system, the next cell of BS-$1$ is $2$, the next cell of BS-$2$ is $3$ and the next cell of BS-$L$ is $1$.} to the $l$th BS are grouped to align the ICI from it in the same interference space. Hence, ICI from the $l$th BS for the users in the next cell span the same subspace as

\setcounter{equation}{9}
\begin{IEEEeqnarray}{rCl}
 {\bf G}_{l} &=& \text{span}\{{\bf H}_{1}^{[l+1,l]H}{\bf U}_{1}^{[l+1]}\} = \text{span}\{{\bf H}_{2}^{[l+1,l]H}{\bf U}_{2}^{[l+1]}\} = \nonumber\\
 && \quad\hdots =  \text{span}\{{\bf H}_{K}^{[l+1,l]H}{\bf U}_{K}^{[l+1]}\}
 \label{eqn:rec_span}
\end{IEEEeqnarray}
\setcounter{equation}{12}

\noindent where span$({\bf A})$ denotes the subspace spanned by the column vectors of any matrix ${\bf A}$. The intersection subspace ${\bf G}_{l}$ of all the ICI and receive beamforming matrices ${\bf U}_{k}^{[l+1]}$ can now be determined by solving the matrix equation (\ref{eqn:U_mat}). The matrix ${\bf X}_{l}$ lies in the null space of $KM\times (M+KN)$ matrix ${\bf F}_{l}$. Therefore, by rank-nullity theorem, minimum receive antennas required for the null space to satisfy the dimensional requirement of the receive beamforming matrices (column dimension must be at least $d_{s}$) are $\tfrac{K-1}{K}M + \tfrac{d_{s}}{K}$. The null space computation can be complex for large matrix sizes, so by utilizing the sparsity of ${\bf F}_{l}$ the authors of \cite{tang} provided a recursive method based on intersection of null spaces to compute the required matrices with lower computation complexity.

The BS-$l$ sees the users of the next cell as a single user due to grouping. Thus by treating the ICI channels corresponding to the users in the next cell as a single ICI channel, the precoding matrices at the $l$th BS can be designed as (\ref{eqn:V_mat}) to communicate with users in its own cell without any interference. Since the size of matrix in (\ref{eqn:V_mat}) is $[K(L-1)d_{s}]\times M $, therefore, the minimum number of transmit antennas required for the null space to have at least $d_{s}$ dimensions are $[K(L-1)+1]\times d_{s}$.

%
%--------------------------------------------------------------------------------
%----------------------------TABLE-I : O-ALGORITHM------------------------------- %--------------------------------------------------------------------------------
%
\begin{center}
\begin{table*}
\centering
\small
\caption{Orthogonality based linear search algorithm}
\begin{tabular}{ll}
\\
1) & \textbf{Initialization}:~Define $\mathcal{T}^{[l]} = \{1,...,K_{l}\}$ for each $1\le l\le L$, initialize the user subsets as \\
&$\mathcal{S}^{[l]} = \underset{K}{\operatorname{arglist}}\max\limits_{j \in \mathcal{T}^{[l]}} ||{\bf H}_{j}^{[l,l]}||_{F}$ for each $1\le l \le L$ such that $\mathcal{S}^{[l]} = \{s_{1}^{l},...,s_{K}^{l}\} ;~C = 0$. Perform the grouping\\
&and compute the initial value of intersection subspace and receiver matrices  ${\bf G}_{l}, {\bf U}_{i}^{[l]}\quad \forall i\in\mathcal{S}^{[l]}, \forall l$\\ [2ex]
2) & for $l = 1:L$ \\
& \quad for $k = 1:K$ \\
& \qquad For every $j \in \mathcal{T}^{[l]}- \{s_{1}^{l},...,s_{k-1}^{l},s_{k+1}^{l},...,s_{K}^{l} \}$, \\[1ex]
& \qquad i) define $\mathcal{S}_{k,j}^{[l]temp} = \{\mathcal{S}^{[l]} | s_{k}^{l} = j\}$.\\[1ex]
& \qquad ii) Compute the temporary intersection subspace and receiver matrix for the users in $\mathcal{S}_{k,j}^{[l]temp}$ using grouping \\
& \qquad \hspace*{15pt} as ${\bf G}_{l}^{temp}$ and ${\bf U}_{i}^{[l]temp},\forall i \in \mathcal{S}_{k,j}^{[l]temp}$. \\[1ex]
& \qquad iii) Compute generator matrix for the desired signal space as ${\bf A}_{G} = [{\bf H}_{j}^{[l,l]H}{\bf U}_{j}^{[l]}]_{o}$ and for the interference space\\
& \qquad \hspace*{15pt} as ${\bf B}_{G} = [{\bf G}_{l}^{temp} \quad
{\bf H}_{t(t\in \mathcal{S}_{k,j}^{[l]temp} - \{j\})}^{[l,l]H}{\bf U}_{t(t\in \mathcal{S}_{k,j}^{[l]temp} - \{j\})}^{[l]temp} \quad {\bf H}_{t(t \in \mathcal{S}^{[m]})}^{[l,m(m = 1,...,L, \neq l,\neq l+1)]H}{\bf U}_{t(t \in \mathcal{S}^{[m]})}^{[m(m = 1,...,L, \neq l,\neq l+1)]} ]_{o}$\\[2ex]
& \qquad $p = \text{arg}\max\limits_{j \in \mathcal{T}^{[l]}- \{s_{1}^{l},...,s_{k-1}^{l},s_{k+1}^{l},...,s_{K}^{l} \}} ||{\bf A}_{G}{\bf A}_{G}^{H} - {\bf B}_{G}{\bf B}_{G}^{H}||_{F}$\\ [1ex]
& \qquad Compute the sum rate as $\mathcal{R}_{p} = \mathcal{R}\left(\mathcal{S}^{[1]},...,\mathcal{S}^{[l-1]},\mathcal{S}_{k,p}^{[l]temp},\mathcal{S}^{[l+1]},...,\mathcal{S}^{[L]}\right)$\\
& \qquad if $\mathcal{R}_{p} > C,$\\ [1ex]
&\qquad \quad $C \leftarrow \mathcal{R}_{p};$\\
&\qquad \quad ${\bf G}_{l} \leftarrow {\bf G}_{l}^{temp}$ and ${\bf U}_{i}^{[l]} \leftarrow {\bf U}_{i}^{[l]temp},\forall i\in \mathcal{S}_{k,p}^{[l]temp};$\\
&\qquad \quad $\mathcal{S}^{[l]} \leftarrow \mathcal{S}_{k,p}^{[l]temp};$\\

\end{tabular}
\label{tab:o_algo}
\vspace*{-5pt}
\end{table*}
\end{center}
%-------------------------------------------------------------------------------- 
%
%--------------------------------------------------------------------------------
%----------------------------SEC-III : USER SELECTION---------------------------- %--------------------------------------------------------------------------------
% 
\section{User Selection}
\label{sec:user_sel}

Suppose the number of users in the $l$th cell is $K_{l}$ and the system can support $K$ users in each cell such that $K < K_{l}, \forall l = 1,...,L$. The sum rate of the system will improve if we utilize multiuser diversity by selecting the optimal user subset from all possible user subsets. Let $\mathcal{T}^{[l]} = \{1,...,K_{l}\}$ denote the set of users in the $l$th cell, $\mathcal{S}^{[l]}$ be the subset of selected users in the $l$th cell and $|\mathcal{S}^{[l]}| = K$, where $|\mathcal{S}^{[l]}|$ denotes the cardinality of $\mathcal{S}^{[l]}$. The sum rate of the system when applied to the selected user subsets $\mathcal{S}^{[l]}, \forall l=1,...,L$ is

\begin{equation}
 \mathcal{R}\left(\mathcal{S}^{[1]},...,\mathcal{S}^{[L]}\right) = \sum\limits_{l = 1}^L\sum\limits_{k \in \mathcal{S}^{[l]}}\mathcal{R}_{k}^{[l]}
 \label{eqn:sum_rate}
\end{equation}

\noindent where (\ref{eqn:sum_rate}) is written using (\ref{eqn:water_fill}) with the users index $k = 1,...,K$ been replaced by the elements of $\mathcal{S}^{[l]}, \forall l$. Hence the maximum sum rate that can be achieved in this MIMO-IFBC is written as

\begin{equation}
 \mathcal{R}_{opt} = \max\limits_{\mathcal{S}^{[l]} \subset \mathcal{T}^{[l]},|\mathcal{S}^{[l]}| = K,  \forall l}\mathcal{R}\left(\mathcal{S}^{[1]},...,\mathcal{S}^{[L]}\right)
 \label{eqn:R_opt}
\end{equation}

\noindent and the $L$ user subsets giving this maximum sum rate, together are optimal user subsets. The total number of brute-force searches to be made to get optimal user subset is $\prod\nolimits_{l = 1}^{L}\binom{K_{l}}{K}$. For e.g., if we take $K = 2, L = 3$ and $K_{l} = 50, \forall l$, then the total number of searches to be made are $1.838\times 10^{9}$ which is quite large even for this small number of users in each cell.

\subsection{Orthogonality approach}
\label{subsec:o_algo}

The brute-force selection algorithm performs search over all possible user subsets and hence the complexity of user selection can be reduced if the search range can be reduced efficiently. In this section we propose a reduced search range suboptimal algorithm such that the complexity of user selection varies linearly with the number of users in each cell. Subsequently, a new user selection metric is proposed to further reduce the complexity with a little compromise in the sum rate performance.

Several user selection algorithms \cite{shen, zhang, lee, tran, gg} exist for single BS MU-MIMO Broadcast channels (BC). However, these algorithms cannot be extended directly to our IFBC with IA. This is because in BC, block diagonalization or other precoding schemes like zero-forcing beamforming are used to completely eliminate the interference received but in IFBC, IA is used to align the interference rather than suppress it completely. Hence in terms of effective downlink channel and received interference, the BC case with orthogonal precoding is fundamentally different from the IFBC case with IA.

In IFBC, to select a user with better effective channel $\bar{{\bf H}}_{k}^{[l,l]}$ we need to compute both ${\bf U}_{k}^{[l]}$ and ${\bf V}_{k}^{[l]}$. To reduce complexity during the user selection process, we aim to eliminate the computation of ${\bf V}_{k}^{[l]}$  which can be done if we know how ${\bf V}_{k}^{[l]}$ is affecting the matrix ${\bf U}_{k}^{[l]H}{\bf H}_{k}^{[l,l]}$. So far we know that ${\bf U}_{k}^{[l]}$ is designed to align the interference channels' space to a common subspace. It is evident from (\ref{eqn:rec_span}) that the interference channels' space can be aligned if the effective downlink channel is of the form ${\bf H}^{H}{\bf U}$. Since we are dealing with received interference, the effective downlink channel will be defined from the viewpoint of the user. In the current scenario, the effective downlink channel is of the form ${\bf H}{\bf V}$, therefore, we need to formulate our problem in a way where the downlink channel matrix has the structure of ${\bf H}^{H}$ and the transmit beamformer could be expressed as a function of ${\bf U}_{k}^{[l]}$. We will address this problem by using the notion of network reciprocity and then formulate the algorithm.
%
%--------------------------------------------------------------------------------
%----------------------------FIG-II : RECIPROCAL SYSTEM-------------------------- %--------------------------------------------------------------------------------
%
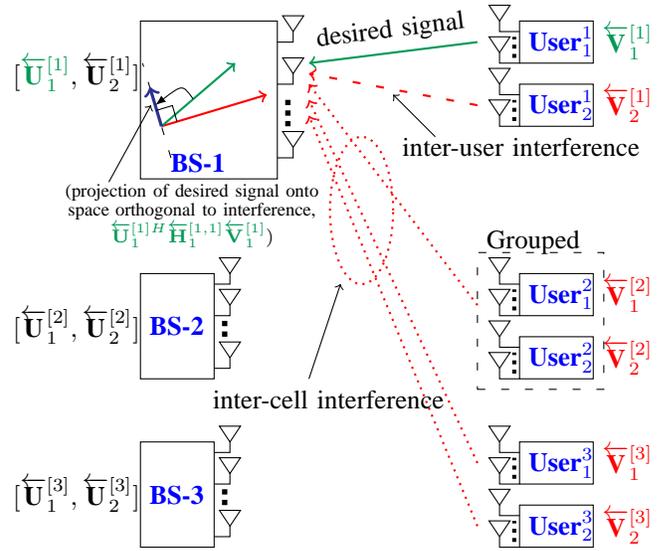
\begin{figure}
\centering
\begin{tikzpicture}[scale = 1.4]

\draw (0.4,0) rectangle (1.1,1);
\node [anchor = west, color=blue] (BS3) at (0.4,0.5) {$\textbf{BS-3}$};
\draw (1.1,0.1) \rightantenna;
\draw (1.1,0.6) \rightantenna;
\draw (1.1,0.9) \rightantenna;
\draw[dotted, line width=2pt] (1.2,0.4) -- (1.2,0.6);
\node [anchor = west] (BS3_V) at (-0.9,0.5) {$[\overleftarrow{\textbf{U}}_{1}^{[3]},\overleftarrow{\textbf{U}}_{2}^{[3]}]$};

\draw (0.4,1.6) rectangle (1.1,2.6) ;
\node [anchor = west, color=blue] (BS2) at (0.4,2.1) {$\textbf{BS-2}$};
\draw (1.1,1.7) \rightantenna;
\draw (1.1,2.2) \rightantenna;
\draw (1.1,2.5) \rightantenna;
\draw[dotted, line width=2pt] (1.2,2.0) -- (1.2,2.2);
\node [anchor = west] (BS2_V) at (-0.9,2.1) {$[\overleftarrow{\textbf{U}}_{1}^{[2]},\overleftarrow{\textbf{U}}_{2}^{[2]}]$};

\draw (0.4,3.5) rectangle (1.7,5) ;
\node [anchor = west, color=blue] (BS1) at (0.6,3.65) {$\textbf{BS-1}$};
\draw (1.7,3.7) \rightantenna;
\draw (1.7,4.4) \rightantenna;
\draw (1.7,4.8) \rightantenna;
 \draw[dotted, line width=2pt] (1.8,4) -- (1.8,4.3);
\node [anchor = west] (BS1_V) at (-0.9,4.5) {$[{\color{ForestGreen}\overleftarrow{\textbf{U}}_{1}^{[1]}},\overleftarrow{\textbf{U}}_{2}^{[1]}]$};

\draw[->, line width=0.8pt, color = red] (0.6,4) -- (1.6,4.3);
\draw[->, line width=0.8pt, color = ForestGreen] (0.6,4) -- (1.3,4.6);
\draw[loosely dashed] (0.69,3.7) -- (0.42,4.6);
\draw[->, line width=1.2pt, color = Blue] (0.6,4) -- (.489,4.372);
\node (A) at (1.6,4.3) {};
\node (B) at (0.6,4) {};
\node (C) at (0.489,4.372) {};
\draw [right angle symbol={A}{B}{C}];
\draw[->, -latex] (0.9,4.26) .. controls (0.8,4.4) .. (.54,4.2);

\node[anchor = west] (h_p) at (-0.4,3.15) {$\displaystyle \substack{\text{(projection of desired signal onto} \\ \text{space orthogonal to interference,} \\ {\color{ForestGreen}{\text{$\overleftarrow{{\bf U}}_{1}^{[1]H}\overleftarrow{{\bf H}}_{1}^{[1,1]}\overleftarrow{{\bf V}}_{1}^{[1]}$}}})}$};
\draw [->] (0.1, 3.47) -- (0.5,4.2);

\draw[ ->, color=ForestGreen, line width=0.8pt] (3.6,4.8) -- (2,4.6) node[pos = 0.5, sloped, above] {{\color{black}desired signal}};
\draw[loosely dashed,->, color=red, line width=0.8pt] (3.6,4.2) -- (2,4.5);
\draw[dotted,->, color=red, line width=0.8pt] (3.6,0.2) -- (2,4.1);
\draw[dotted,->, color=red, line width=0.8pt] (3.6,0.8) -- (2,4.25);
\draw[dotted,->, color=red, line width=0.8pt] (3.6,2.3) -- (2,4.4);

\draw (4,0) rectangle (4.7,0.4) ;
\node [anchor = west, color=blue] (User23) at (4,0.2) {$\textbf{User}_{2}^{3}$};
\draw (4,0) \leftantenna;
\draw (4,0.3) \leftantenna;
\draw[dotted, line width = 1.5pt] (3.95,0.05) -- (3.95,0.25);
\node [anchor = west, color=red] (U23) at (4.7,0.2) {$\overleftarrow{\textbf{V}}_{2}^{[3]}$};

\draw (4,0.6) rectangle (4.7,1) ;
\node [anchor = west, color=blue] (User13) at (4,0.8) {$\textbf{User}_{1}^{3}$};
\draw (4,0.6) \leftantenna;
\draw (4,0.9) \leftantenna;
\draw[dotted, line width = 1.5pt] (3.95,0.65) -- (3.95,0.85);
\node [anchor = west, color=red] (U13) at (4.7,0.8) {$\overleftarrow{\textbf{V}}_{1}^{[3]}$};

\draw (4,1.6) rectangle (4.7,2.0) ;
\node [anchor = west, color=blue] (User22) at (4,1.8) {$\textbf{User}_{2}^{2}$};
\draw (4,1.6) \leftantenna;
\draw (4,1.9) \leftantenna;
\draw[dotted, line width = 1.5pt] (3.95,1.65) -- (3.95,1.85);
\node [anchor = west, color=red] (U22) at (4.7,1.8) {$\overleftarrow{\textbf{V}}_{2}^{[2]}$};

\draw (4,2.2) rectangle (4.7,2.6) ;
\node [anchor = west, color=blue] (User12) at (4,2.4) {$\textbf{User}_{1}^{2}$};
\draw (4,2.2) \leftantenna;
\draw (4,2.5) \leftantenna;
\draw[dotted, line width = 1.5pt] (3.95,2.25) -- (3.95,2.45);
\node [anchor = west, color=red] (U12) at (4.7,2.4) {$\overleftarrow{\textbf{V}}_{1}^{[2]}$};

\draw [loosely dashed] (3.6,1.5) rectangle (4.8,2.8) ;
\node [anchor = west] (Group) at (3.6, 2.9) {Grouped};

\draw (4,4) rectangle (4.7,4.4) ;
\node [anchor = west, color=blue] (User21) at (4,4.2) {$\textbf{User}_{2}^{1}$};
\draw (4,4) \leftantenna;
\draw (4,4.3) \leftantenna;
\draw[dotted, line width = 1.5pt] (3.95,4.05) -- (3.95,4.25);
\node [anchor = west, color=red] (U21) at (4.7,4.2) {$\overleftarrow{\textbf{V}}_{2}^{[1]}$};

\draw (4,4.6) rectangle (4.7,5) ;
\node [anchor = west, color=blue] (User11) at (4,4.8) {$\textbf{User}_{1}^{1}$};
\draw (4,4.6) \leftantenna;
\draw (4,4.9) \leftantenna;
\draw[dotted, line width = 1.5pt] (3.95,4.65) -- (3.95,4.85);
\node [anchor = west, color=ForestGreen] (U11) at (4.7,4.8) {$\overleftarrow{\textbf{V}}_{1}^{[1]}$};

\node [anchor = west] (IUI) at (2.8,3.8) {inter-user interference};
\draw[->] (3.2,3.9) -- (2.8,4.3);
\draw[dotted, line width=0.8pt, color = red] (2.5,3.2) ellipse (0.3 and 0.7);
\node [anchor = west] (ICI) at (1,1.4) {inter-cell interference};
\draw[->] (2,1.6) -- (2.3,2.5);

\end{tikzpicture}
\caption{Reciprocal system of the MIMO-IFBC with $L = 3$ and $K = 2$ in each cell, users in cell-$2$ are grouped and the BS-$1$ is shown to be performing orthogonal projection of the received signal space.}
\label{fig:reci_IFBC}
\end{figure}
%-------------------------------------------------------------------------------- 

\begin{figure*}[!t]
 \normalsize
 \setcounter{MYtempeqncnt}{\value{equation}}
 \setcounter{equation}{17}
 
 \begin{equation}
  \overleftarrow{{\bf U}}_{k}^{[l]} \subset [
   \underbrace{\overleftarrow{{\bf H}}_{t(t=1,...,K)}^{[l,s(s\neq l)]}\overleftarrow{{\bf V}}_{t(t=1,...,K)}^{[s(s\neq l)]}}_{\text{effective ICI channels}} 
   \quad
   \underbrace{\overleftarrow{{\bf H}}_{t(t=1,...,K, \neq k)}^{[l,l]}\overleftarrow{{\bf V}}_{t(t=1,...,K, \neq k)}^{[l]}}_{\text{effective IUI channels}}
   ]^{\perp}
  \label{eqn:U_mat_rec}
 \end{equation}
 \hrulefill
  \begin{equation}
  {\bf V}_{k}^{[l]} \subset [
   \underbrace{{\bf G}_{l}}_{\text{effective interference channels}} \quad
   \underbrace{{\bf H}_{t(t=1,...,K)}^{[s(s\neq l,\neq l+1), l]H}{\bf U}_{t(t=1,...,K)}^{[s(s\neq l,\neq l+1)]}}_{\text{effective ICI channels}} 
   \quad
   \underbrace{{\bf H}_{t(t=1,...,K, \neq k)}^{[l,l]H}{\bf U}_{t(t=1,...,K, \neq k)}^{[l]}}_{\text{effective IUI channels}}
   ]^{\perp}
  \label{eqn:U_mat_rec_next}
 \end{equation}
 \hrulefill
\end{figure*}

%
%--------------------------------------------------------------------------------
%----------------------------TABLE-II : S-ALGORITHM------------------------------ %--------------------------------------------------------------------------------
%
\begin{center}
\begin{table*}
\centering
\small
\caption{Sum rate based linear search algorithm}
\begin{tabular}{ll}
\\
1) & \textbf{Initialization}:~Define $\mathcal{T}^{[l]} = \{1,...,K_{l}\}$ for each $1\le l\le L$, initialize the user subsets as \\
&$\mathcal{S}^{[l]} = \underset{K}{\operatorname{arglist}}\max\limits_{j \in \mathcal{T}^{[l]}} ||{\bf H}_{j}^{[l,l]}||_{F}$ for each $1\le l \le L$ such that $\mathcal{S}^{[l]} = \{s_{1}^{l},...,s_{K}^{l}\} ;~C = 0$. Perform the grouping\\
&and compute the initial value of receiver matrices ${\bf U}_{i}^{[l]}, \forall i \in \mathcal{S}^{[l]}, \forall l$\\ [2ex]
2) & for $l = 1:L$ \\
& \quad for $k = 1:K$ \\
& \qquad For every $j \in \mathcal{T}^{[l]} - \{s_{1}^{l},...,s_{k-1}^{l},s_{k+1}^{l},...,s_{K}^{l} \}$, \\[1ex]
& \qquad i) define $\mathcal{S}_{k,j}^{[l]temp} = \{\mathcal{S}^{[l]} | s_{k}^{l} = j\}$.\\[1ex]
& \qquad ii) Compute the temporary receiver matrix for the users in $\mathcal{S}_{k,j}^{[l]temp}$ using grouping as ${\bf U}_{j}^{[l]temp}$. \\[1ex]
& \qquad iii) Using the ${\bf U}_{j}^{[l]temp}$ and ${\bf U}_{i}^{[m]}, i \in \mathcal{S}^{[m]}~\forall m\neq l$ compute the transmit processing matrices using (\ref{eqn:V_mat}) as ${\bf V}_{i}^{[m]}, i \in \mathcal{S}^{[m]}$\\
&\qquad \hspace{15pt}$~\forall m \neq l\text{ and }i \in \mathcal{S}_{k,j}^{[l]temp}, \text{ for } m = l$.\\[1ex]
&\qquad iv) Using the computed values of receive and transmit matrices compute $\mathcal{R}_{j} = \mathcal{R}\left(\mathcal{S}^{[1]},...,\mathcal{S}^{[l-1]},\mathcal{S}_{k,j}^{[l]temp},\mathcal{S}^{[l+1]},...,\mathcal{S}^{[L]}\right)$ \\
&\qquad \hspace{15pt} using (\ref{eqn:sum_rate}) for the selected users.\\ [1ex]
& \qquad $p = \text{arg}\max\limits_{j \in \mathcal{T}^{[l]}- \{s_{1}^{l},...,s_{k-1}^{l},s_{k+1}^{l},...,s_{K}^{l} \}} \mathcal{R}_{j}$\\ [1ex]
& \qquad if $\mathcal{R}_{p} > C,$\\ [1ex]
&\qquad \quad $C \leftarrow \mathcal{R}_{p};$ \\
&\qquad \quad ${\bf U}_{i}^{[l]} \leftarrow {\bf U}_{i}^{[l]temp}, \forall i \in \mathcal{S}_{k,p}^{[l]temp};$ \\
&\qquad \quad $\mathcal{S}^{[l]} \leftarrow \mathcal{S}_{k,p}^{[l]temp};$\\

\end{tabular}
\label{tab:s_algo}
\end{table*}
\end{center}
%--------------------------------------------------------------------------------

\subsubsection{Network Reciprocity}

The reciprocal channel model in \cite{gomadan, babadi} will be used to exploit network reciprocity concepts. In the reciprocal system, the role of transmitter and receiver are switched. For e.g., a transmitter in the original system becomes the receiver in the reciprocal system. The $M\times N$ channel matrix at the $l$th receiver from the $k$th transmitter in the $j$th cell is denoted by $\overleftarrow{{\bf H}}_{k}^{[l,j]} = {\bf H}_{k}^{[j,l]H}$ in the reciprocal system. Similarly, the $N\times 1$ transmit and $M\times 1$ receive beamforming matrices are $\overleftarrow{{\bf V}}_{k}^{[l]}$ and $\overleftarrow{{\bf U}}_{k}^{[l]}$ respectively. The total transmit power is assumed to be same as in the reciprocal system. The reciprocal system of the IFBC in \figurename\,\ref{fig:IFBC_sys} is shown in \figurename\,\ref{fig:reci_IFBC} where the grouping of users is also performed. The interference channel space from the $\text{User}_{1}^{2}$ and $\text{User}_{2}^{2}$ in cell-$2$ overlap because of grouping and hence BS-$1$ sees them as a single user (refer \ref{eqn:rec_span}). 

The feasibility conditions on the reciprocal system become

\begin{IEEEeqnarray}{rCl}
\setcounter{equation}{15}
&&\overleftarrow{{\bf U}}_{k}^{[l]H}\overleftarrow{{\bf H}}_{k}^{[l,l]}\overleftarrow{{\bf V}}_{i}^{[l]} = 0, \quad \forall i \neq k 
\label{eqn:feas_recip_1}\\
&&\overleftarrow{{\bf U}}_{k}^{[l]H}\overleftarrow{{\bf H}}_{k}^{[l,j]}\overleftarrow{{\bf V}}_{m}^{[j]} = 0, \quad \forall m = 1,...,K \text{ and } \forall j \neq l  \label{eqn:feas_recip_2}\\
&&\text{rank}\left(\overleftarrow{{\bf U}}_{k}^{[l]H}\overleftarrow{{\bf H}}_{k}^{[l,l]}\overleftarrow{{\bf V}}_{k}^{[l]} \right) = d_{s}
\label{eqn:feas_recip_3}
\end{IEEEeqnarray}
\setcounter{equation}{19}

If we set $\overleftarrow{{\bf V}}_{k}^{[l]} = {\bf U}_{k}^{[l]}$ and $\overleftarrow{{\bf U}}_{k}^{[l]} = {\bf V}_{k}^{[l]}$ then the feasibility conditions on the reciprocal system become identical to the feasibility conditions in  (\ref{eqn:rec_reqr_1})$-$(\ref{eqn:rec_reqr_3}). This is called Reciprocity of Alignment \cite{gomadan}. Another implication of the reciprocity of alignment is that any scheme derived in the reciprocal system will work in the original system if the roles of transmit and receive beamformers are reversed.

Now returning to the issue of expressing the effective downlink channel in terms of channel matrix and receive beamformer (to take into account grouping), we can determine the receive beamformer in the reciprocal system as in (\ref{eqn:U_mat_rec}) using (\ref{eqn:feas_recip_1})$-$(\ref{eqn:feas_recip_2}) to completely eliminate the received IUI and ICI. By using the reciprocity of alignment, (\ref{eqn:U_mat_rec}) is written as (\ref{eqn:U_mat_rec_next}) where we have grouped the ICI terms of the next cell using (\ref{eqn:rec_span}) and thus have taken care of grouping in the design of $\overleftarrow{{\bf U}}_{k}^{[l]}$.  

From \figurename\,\ref{fig:reci_IFBC} it can be seen that the user whose desired signal space is close to orthogonal of the received interference space will have better projection onto space orthogonal to interference. This leads to better effective channel $\overleftarrow{{\bf U}}_{k}^{[l]H}\overleftarrow{{\bf H}}_{k}^{[l,l]}\overleftarrow{{\bf V}}_{k}^{[l]}$ which is equal to $\bar{{\bf H}}_{k}^{[l,l]H}$ from (\ref{eqn:eff_channel}). Therefore, to get better effective channel without computing $\overleftarrow{{\bf U}}_{k}^{[l]}$ (or ${\bf V}_{k}^{[l]}$ in the original system), the user whose desired signal space is closest to orthogonal of the received interference space should be selected. The maximum chordal distance criteria can be used to incorporate this orthogonality requirement.

\subsubsection{Chordal distance}

The Grassmannian space $G(m,n)$ is the set of all $n$-dimensional subspaces of Euclidean $m$-dimensional space \cite{grass}. A $m\times n$ matrix is called the generator matrix for an $n$-plane $P \in G(m,n)$ if its columns span $P$. Suppose ${\bf A}_{G}$ and ${\bf B}_{G}$ are generator matrices of planes $P$ and $Q$,  columns of which are orthonormal vectors, then the chordal distance between $P$ and $Q$ is defined as

\begin{equation}
 d_{c}\left(P, Q\right) = \frac{1}{\sqrt{2}}||{\bf A}_{G}{\bf A}_{G}^{H} - {\bf B}_{G}{\bf B}_{G}^{H}||_{F}
 \label{eqn:cho_dist}
\end{equation}

Chordal distance is known to be proportional to the degree of orthogonality between the subspaces.

\subsubsection{User Selection Algorithm}

By using the above concepts as building blocks the user selection algorithm is formulated as follows. For initializing the algorithm, we will rank the users on the basis of their channel energy (channel frobenius norm). Therefore, $K$ users are selected in each cell with maximum channel frobenius norm. The receiver processing matrices ${\bf U}_{k}^{[l]}$ are then computed for the users initialized in each cell. We compute ${\bf U}_{k}^{[l]}$ using (\ref{eqn:U_mat}) if $K \le 3$ and use the decoupled approach \cite{tang} if $K > 3$. Let us define $\mathcal{S}_{k,j}^{[l]temp}$ as a temporary user subset whose elements are same as those of $\mathcal{S}^{[l]}$ except for the $k$th element $s_{k}^{l}$ which is replaced by the element $j$. The algorithm proceeds by employing coordinate ascent approach \cite{ascent}. At each step generator matrices for the desired signal space and the interference space in the reciprocal system are computed. Let the columns of ${\bf A}_{o}$  be the orthonormal basis of the column space of matrix ${\bf A}$. The matrix ${\bf A}_{o}$ can be computed by applying the Gram-Schmidt Orthogonalization (GSO) procedure to the columns of ${\bf A}$. At each step, the user with maximum chordal distance is selected from the remaining users. The selected user with maximum chordal distance will replace the existing user in the initialized user subset only if the sum rate on its involvement increases. The orthogonality based user selection algorithm is summarized in Table-\ref{tab:o_algo} where $\underset{K}{\operatorname{arglist}}$ in the initialization step gives as output a list of arguments of length $K$.

It may be noted that in the above algorithm the BSs share information to perform user selection in a distributed manner. Thus, the $l$th BS $(0 \leq l \leq L)$ needs $\mathcal{S}^{[m]} ,~ m \neq l$ (most recent of either initialized $\mathcal{S}^{[m]}$ in Step-$1$ or $\mathcal{S}^{[m]}$ updated by the $m$th BS in Step-$2$)  before initializing user selection in its own cell.

\subsection{Sum rate approach}
\label{subsec:s_algo}

The user selection algorithm proposed in \cite{inkyu} for IFC can be extended to the IFBC case because it is directly utilizing sum rate as the selection criteria. However, the direct extension of the algorithm by computing ${\bf U}_{k}^{[l]}$ and ${\bf V}_{k}^{[l]}$ in each iteration to evaluate the sum rate is not computationally ideal. For e.g., in the step where users in cell-$1$ are being updated the index of the users already selected in other cells are fixed, so in each iteration the ${\bf U}_{k}^{[l]}$ matrix needs to be computed only for the users in cell-$1$ and ${\bf U}_{k}^{[l]}$ will remain same for users in rest of the cells.

The extension of the algorithm is formulated as follows. The algorithm is initialized similar to the previous algorithm. To save the unnecessary computation of ${\bf U}_{k}^{[l]}$ we will be updating these matrices in the similar fashion as the previous algorithm. Using coordinate ascent approach \cite{ascent} the user subsets are updated such that sum rate is maximized in each step. The user index $s_{k}^{l}$ is varied over all remaining users index in the $l$th cell $\mathcal{T}^{[l]} - \{s_{1}^{l},...,s_{k-1}^{l},s_{k+1}^{l},...,s_{K}^{l} \}$, to find the one giving maximum sum rate, keeping index of the other selected users unchanged. A user will replace the existing user in the user subset only if it increases the sum rate. This procedure is performed for each user index in each of the $L$ user subsets. The sum rate approach based user selection algorithm is summarized in Table-\ref{tab:s_algo}.

Though the proposed algorithms have been shown to work with the extended grouping scheme, they could be easily extended to other existing IA schemes like the one proposed in \cite{ma}. This scheme is good in the sense that it requires less number of antennas as compared to the extended grouping scheme. However, it should be noted that this scheme is applicable only when $L = 3$.
%
%--------------------------------------------------------------------------------
%-------------------------SEC-IV : COMPLEXITY ANALYSIS---------------------------
%--------------------------------------------------------------------------------
%
\section{Complexity Analysis}
\label{sec:comp_anal}

In this section the computation complexity of the algorithms is discussed using flop count. The complexity of an operation is counted as total number of flops required which we denote as $\psi$. A flop is defined as a real floating point operation \cite{golub}. A real multiplication, addition is counted as one flop and hence a complex multiplication and addition will count as six and two flops respectively. For simplicity we will assume that the number of users in each cell $K_{l} = K_{T}$. We discuss the flop count of some typical matrix operations \cite{shen, hunger} for a complex valued $N\times M$ matrix ${\bf H}$ as follows. The computation of frobenius norm of ${\bf H}$ requires $4MN$ flops, GSO$({\bf H})$ takes $8N^2M - 2MN$ flops and the approximate flops required to compute Singular Value Decomposition (SVD) of ${\bf H}$ are

\begin{equation}
 \psi_{SVD}(N, M) = 24NM^2 + 48N^2M + 54N^3
 \label{eqn:flop_SVD}
\end{equation}

\subsection{Orthogonality Approach}
\label{subsec:comp_cho}

The initialization of the algorithm in Table-\ref{tab:o_algo} requires $K_{T}\times L$ frobenius norm computations, hence flops required are $K_{T}L\times 4MN$. Let $\psi_{U}$ denote the flops required to compute the receiver beamforming matrix ${\bf U}_{k}^{[l]}$ for all users in the $l$th cell. The computation of ${\bf U}_{k}^{[l]}$ from (\ref{eqn:U_mat}) requires SVD computation of $KM\times [M+KN]$ matrix, hence $\psi_{U} = \psi_{SVD}(KM, M + KN)$. However, in \cite{tang} a decoupled approach is proposed to reduce the complexity of computation of receive beamformer utilizing intersection of the null spaces, dimension of which reduces with each recursion. The method is effective in complexity reduction for $K > 3$ and the flops required are

\begin{IEEEeqnarray}{rCl}
\psi_{U} &=& K\times \psi_{SVD}(M, M+N) + \sum\limits_{i = 1}^{\lceil\text{log}_{2}K\rceil}\left\{\left\lceil\frac{K}{2^{i}}\right\rceil\times\right.\nonumber\\
&&\quad\left.\left(\psi_{SVD}(M, 2^{i-1}N - s_{i}M) \right.\right.\nonumber\\
&& \quad\left.\left.{+}\: 8M(2^{i-1}N - s_{i}M)(2^{i}N - s_{i+1}M)\right)\right.\nonumber\\
&&\left.{+}\: K\times8N(2^{i-1}N - s_{i}M)(2^{i}N - s_{i+1}M)\vphantom{\frac{1}{2}}\right\}
\label{eqn:flop_U_dec}
\end{IEEEeqnarray}

\noindent where $s_{1} = 0, s_{i} = 2s_{i-1} + 1$ and $\lceil a\rceil$ is the smallest integer number greater than or equal to $a$. The computation of generator matrices ${\bf A}_{G}$ and ${\bf B}_{G}$ involve matrix multiplication requiring $8MNd_{s}$ and $[K(L-1)-1]\times 8MNd_{s}$ flops, respectively and GSO procedure requiring $8M^2d_{s}-2Md_{s}$ and $8M^2(K(L-1)d_{s}) - 2M(K(L-1)d_{s})$ flops respectively. For matrix product ${\bf A}_{G}{\bf A}_{G}^{H}$ and ${\bf B}_{G}{\bf B}_{G}^{H}$, flops required are $8M^2d_{s}$ and $8M^2(K(L-1)d_{s})$. The frobenius norm of $\left({\bf A}_{G}{\bf A}_{G}^{H} - {\bf B}_{G}{\bf B}_{G}^{H}\right)$ requires $6M^2$ flops. The flops required to compute the sum rate $\mathcal{R}_{p}$ are ignored. The total flops for the algorithm are

\begin{eqnarray}
\psi_{cho} &\approx &4K_{T}LMN + L\psi_{U} + \left\{ \psi_{U} + 8M^2d_{s} - 2Md_{s}\right.\nonumber \\
&& \left.\quad{+}\: 8MNd_{s}\times [K(L-1)]\right.\nonumber\\
&& \left.\quad{+}\: 8M^2(K(L-1)d_{s}) - 2M(K(L-1)d_{s})\right.\nonumber \\
&& \left.\quad{+}\:  8M^2d_{s} + 8M^2(K(L-1)d_{s}) + 6M^2\vphantom{\psi_{U} + M^2}\right\}\nonumber \\
&&\times (K_{T} - K +1)KL
\label{eqn:flop_o_algo}
\end{eqnarray}

\noindent and hence complexity of the algorithm varies linearly with the number of users in each cell $(K_{T})$.

\subsection{Sum rate approach}
\label{subsec:comp_c_algo}

The flops required in initialization in Table-\ref{tab:s_algo} are similar to previous algorithm, $K_{T}L\times 4MN$. The flops required to compute the receive beamforming matrices in a particular cell are $\psi_{U}$, like in the previous algorithm. The transmit matrix for the $k$th user in the $l$th cell, ${\bf V}_{k}^{[l]}$ needs SVD computation of $M\times [K(L-1)\times d_{s}]$ matrix, hence flops required are $\psi_{SVD}(M, K(L-1)\times d_{s})$. To compute the pre-whitening filter ${\bf W}_{k}^{[l]}$, $8d_{s}^2N$ flops are required for matrix multiplication. The complexity of inverse of $d_{s}\times d_{s}$ matrix is ignored. The computation of sum rate using (\ref{eqn:water_fill}) involves the multiplication ${\bf W}_{k}^{[l]}\bar{{\bf H}}_{k}^{[l,l]}$, complexity of which is $8NMd_{s} + 8Md_{s}^{2} + 8d_{s}^{3}$. The flops required by the water-filling over $d_{s}$ eigenmodes are ignored since $d_{s}$ is smaller than $M$ and $N$. Therefore, the total flops of the algorithm are

\begin{eqnarray}
\psi_{s} &\approx & 4K_{T}LMN + L\psi_{U} + \left\{\psi_{U} \vphantom{d_{s}^2}\right.\nonumber\\
&& \left.\quad{+}\: KL\times\left[\psi_{SVD}(M,K(L-1)d_{s})\vphantom{d_{s}^2}\right.\right. \nonumber \\
&& \left.\left.\qquad{+}\: (8d_{s}^2N + 8NMd_{s} + 8Md_{s}^{2} + 8d_{s}^{3})\right]\right\} \nonumber \\
&&\times (K_{T}-K+1)KL
\label{eqn:flop_s_algo}
\end{eqnarray}

\subsection{Brute-force Approach}
\label{subsec:comp_opt}

The flop count for brute-force selection algorithm to obtain the optimal solution can be written as

\begin{eqnarray}
\psi_{opt} &\approx &\left[\binom{K_{T}}{K}\right]^{L}\times \left\{ KL\times\psi_{SVD}(M,K(L-1)d_{s})\vphantom{d_{s}^2}\right. \nonumber\\
&& \left.\quad{+}\:  L\psi_{U} + KL\times\right.\nonumber \\
&&\left.\qquad (8d_{s}^2N + 8NMd_{s} + 8Md_{s}^2 + 8d_{s}^3)\right\}\nonumber \\
&\approx & \mathcal{O}\left(K_{T}^{KL}K^{-KL-\tfrac{L}{2}+1}M^{3}L \right)
\label{eqn:flops_opt}
\end{eqnarray}

\noindent where the flops count $\psi_{U}$ is determined for $K \le 3$ as an example to demonstrate the complexity order. The order is shown to be exponential in $K_{T}$ and we have used the Stirling's approximation \cite{knuth} to the factorial and approximated the binomial coefficient as 

\begin{equation}
\binom{K_{T}}{K} \approx K_{T}^{K}K^{-K-\tfrac{1}{2}}\nonumber
\end{equation}
% %--------------------------------------------------------------------------------
%--------------------------SEC-V : SIMULATION RESULTS----------------------------
%--------------------------------------------------------------------------------
% 
\section{Simulation Results}
\label{sec:sim}

In this section, we provide the sum rate and flop count results for the orthogonality approach (o-algorithm) and sum rate approach (s-algorithm) and compare them with the brute-force selection algorithm. The sum rate results are averaged over $1000$ random channel realizations. We will assume that the number of users in each cell $K_{l} = K_{T}, \forall l$. The total transmit power of each BS is fixed at $P$ i.e. $P_{l} = P, \forall l$. The simulation results are shown for different values of total transmit power to noise variance ratio $(\text{SNR} = \tfrac{P}{\sigma^2})$ in dB.

\begin{figure}
\centering
\includegraphics*[viewport=395 2 810 355, width = 3.2in, height = 2.7in]{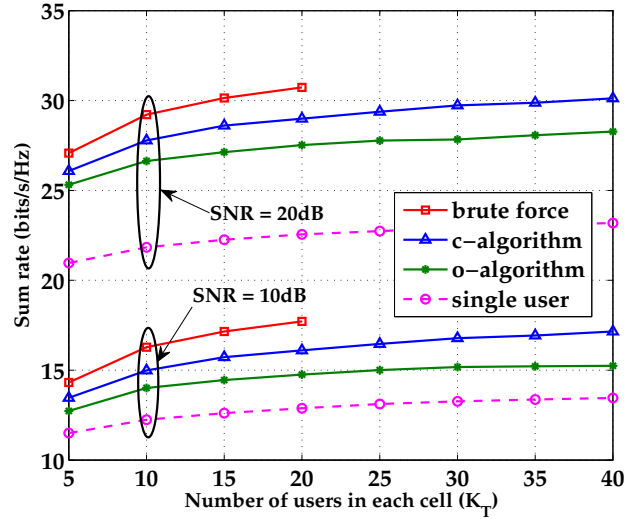}
\caption{Sum rate versus number of users in each cell when $M = 3, N = 2, K = 2, L = 2$ and $d_{s} = 1$.}
\label{fig:per_3_2}
\end{figure}

\begin{figure}
\centering
\includegraphics*[viewport=395 2 810 355, width = 3.2in, height = 2.7in]{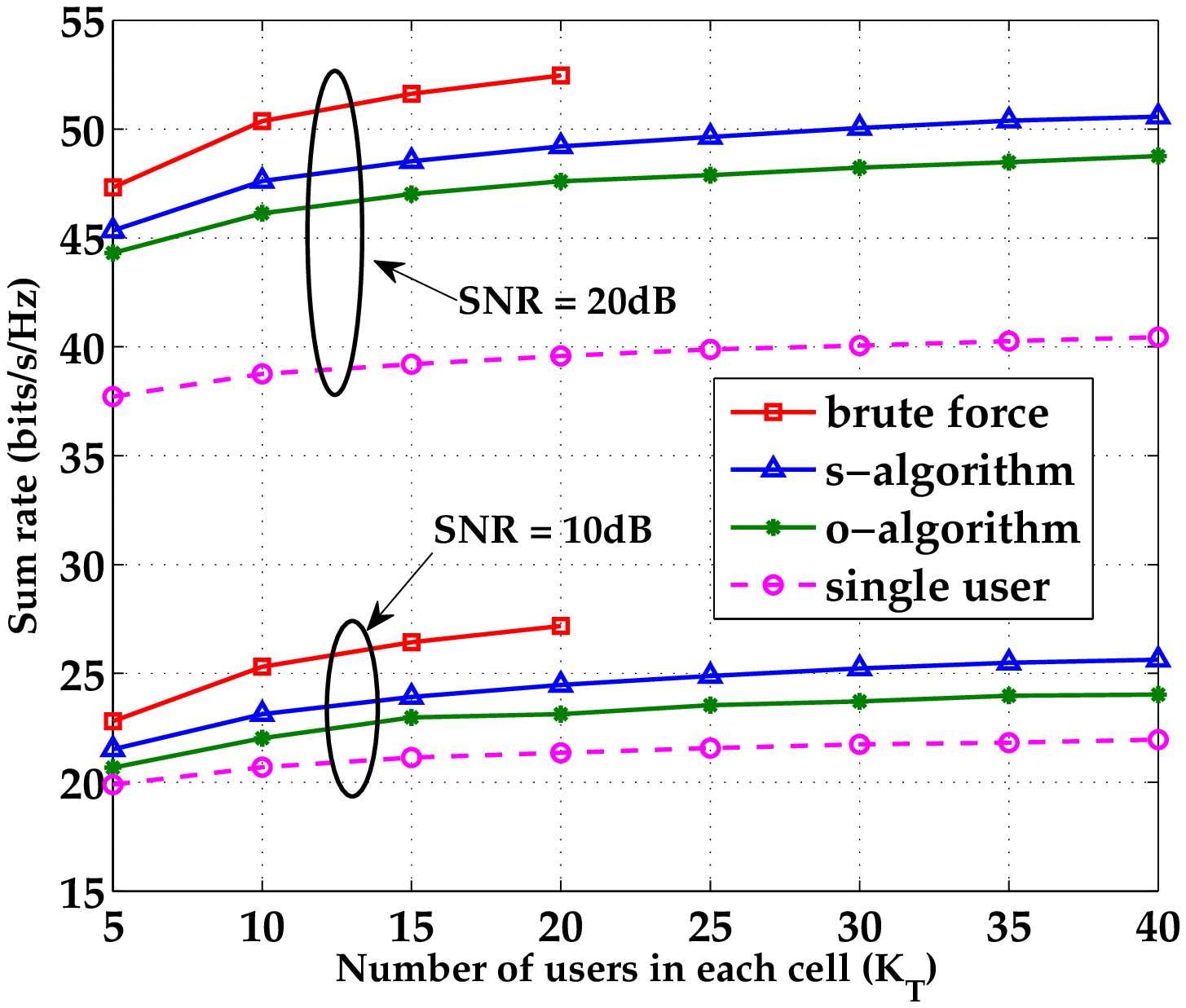}
\caption{Sum rate versus number of users in each cell when $M = 6, N = 4, K = 2, L = 2$ and $d_{s} = 2$.}
\label{fig:per_6_4}
\end{figure}

\begin{figure}
\centering
\includegraphics*[viewport=395 2 810 375, width = 3.2in, height = 2.7in]{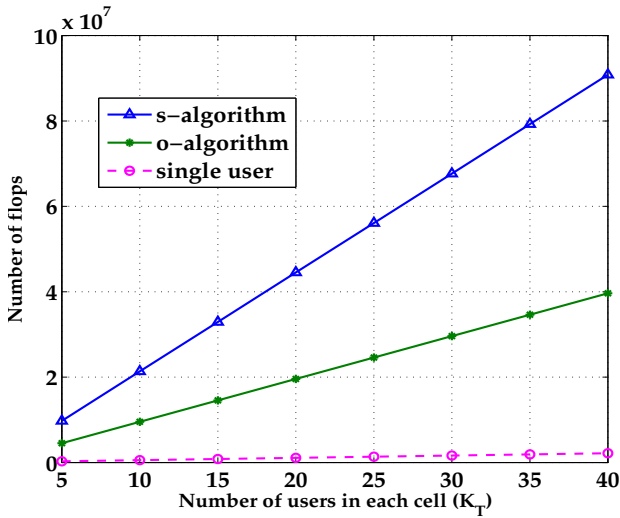}
\caption{Number of flops versus number of users in each cell when $M = 6, N = 4, K = 2, L = 2$ and $d_{s} = 2$.}
\label{fig:flops_6_4}
\end{figure}

In \figurename\,\ref{fig:per_3_2} and \figurename\,\ref{fig:per_6_4} the sum rate is compared for various user selection algorithms with respect to the number of users in each cell $(K_{T})$ for two values of SNR, $10\text{~dB}$ and $20\text{~dB}$. It can be observed that the sum rate achieved by the two suboptimal algorithms namely s-algorithm and o-algorithm is more than $90\%$ of the optimal sum rate achieved by the brute-force selection algorithm. The reduction in achievable sum rate in these suboptimal algorithms is because the search range of users is reduced. However, this reduction in search range has a  significant impact on complexity. Thus, as we can see from (\ref{eqn:flops_opt}), the complexity of brute-force search is exponential with respect to $K_{T}$ as compared to linear for the above suboptimal algorithms. Whenever the same search method is used (coordinate ascent approach here), the sum rate achieved by the s-algorithm is higher as compared to the o-algorithm because it directly uses the sum rate as selection metric. It may be noted that the sum rate achieved by the o-algorithm is close to that achieved by the s-algorithm because it inherently takes care of grouping by using the notion of reciprocal system and uses orthogonality criterion to select the user with better effective channel.

Further, in \figurename\,\ref{fig:per_3_2} and \figurename\,\ref{fig:per_6_4} we also plot the existing algorithm \cite{inkyu} which selects a single user in each cell in IFC. To achieve optimal dof in a two cell IFC, a scheme proposed by the authors of \cite{fakher} already exists in the literature. Therefore, we have used the user selection algorithm of \cite{inkyu} in the system model of \cite{fakher}. The achievable dof in IFBC in \figurename\,\ref{fig:per_3_2} (\figurename\,\ref{fig:per_6_4}) are $d_{s} = 1~(2)$ for each user making a total of $4~(8)$ dof while in IFC the total is min$\{2M, 2N, \text{max}(M,N)\}$ \cite{fakher} which is equal to $3~(6)$ dof. This explains the significant improvement in the sum rate when multiple users are selected than single user selection.

In \figurename\,\ref{fig:flops_6_4} the flop count of the two suboptimal algorithms for multi-user selection and of the algorithm \cite{inkyu} for single user selection is compared as a function of the number of users in each cell $(K_{T})$. Since $K\le 3$, the ${\bf U}_{k}^{[l]}$ is computed using (\ref{eqn:U_mat}) and the flop count $\psi_{U}$ will be used in (\ref{eqn:flop_o_algo}), (\ref{eqn:flop_s_algo}) accordingly. It can be seen that the total flop count of the o-algorithm is nearly half of the total flop count of s-algorithm. The reduction in complexity is because the sum rate computation in each step of the s-algorithm requires two SVD computation, one for ${\bf U}_{k}^{[l]}$ and other for ${\bf V}_{k}^{[l]}$, however, in o-algorithm the computation of ${\bf V}_{k}^{[l]}$ is not required. The computation of chordal distance is much less complex as compared to SVD computation, and this computation gain increases with increase in number of antennas.

It can be observed from \figurename\,\ref{fig:per_3_2} and \figurename\,\ref{fig:per_6_4} that the difference between the sum rate achieved by the s-algorithm and the o-algorithm becomes nearly constant as $K_{T}$ increases. However, from \figurename\,\ref{fig:flops_6_4} we can see that difference between the flop count of these algorithms increase with $K_{T}$. So o-algorithm is preferable when the number of users in each cell is large.
% %--------------------------------------------------------------------------------
%--------------------------SEC-VI : CONCLUSIONS----------------------------------
%--------------------------------------------------------------------------------
%
\section{Conclusions}
\label{sec:concl}

The user selection problem has been addressed to improve the achievable sum rate of the MIMO-IFBC system. A suboptimal user selection algorithm is proposed to reduce the complexity of selection process. The algorithm exploits network reciprocity concepts and orthogonality between the desired signal space and interference space in the reciprocal system to select the users. An existing suboptimal algorithm based on the sum rate criteria is also extended to MIMO-IFBC. Simulation results show that the sum rate achieved by the orthogonality based algorithm and the extended sum rate based algorithm is close to the optimal sum rate. The complexity of these algorithms turns out to be linear with respect to the number of users in each cell as compared to exponential for brute-force search.
%
%--------------------------------------------------------------------------------
%--------------------------------REFERENCES--------------------------------------
%--------------------------------------------------------------------------------
%
\bibliographystyle{IEEEtran}
\bibliography{IEEEabrv,MIMO_IFBC}
%
%--------------------------------END DOCUMENT------------------------------------
%
\end{document}